\documentclass[12pt,dvips]{article}

\usepackage{graphicx}
\usepackage{latexsym}
\usepackage{epsfig}

\newcommand{\bra}[1]{\langle #1|}
\newcommand{\ket}[1]{|#1\rangle}
\newcommand{\braket}[2]{\langle #1|#2\rangle}

\begin{document}

\thispagestyle{empty}

\date{\today}
\title{
\vspace*{2.5cm}
Dirac zero-modes in compact $U(1)$ gauge theory }
\author
{\bf Thomas Drescher and C.B. Lang  \\  \\
Institut f\"ur Physik (Theoretische Physik),\\
Universit\"at Graz, A-8010 Graz, AUSTRIA}
\maketitle
\begin{abstract}
We study properties of the zero and near-zero eigenmodes of the overlap Dirac operator in compact
$U(1)$ gauge theory. In the confinement phase the exact zero-modes are localized as
found by studying the values of the inverse participation ratio
and other features. Non-zero-eigenmodes are less localized in the confinement phase.
In the Coulomb phase no zero-modes are observed and the eigenmodes show
no localization at all.
\end{abstract}



\newpage
\section{Motivation}

The Atiyah-Singer index theorem \cite{AtSi71} relates the topological charge of
the background gauge field configuration to the number of fermionic zero-modes
of the Dirac operator. The theorem states a particular connection between the 
background gauge field and the fermionic fields.  It is a theorem derived on
differentiable manifolds, however. Quantum field configurations and in
particular lattice fields are  non-differentiable. Hence it is of interest to
study on the lattice the topology of the gauge background as well as the
properties of the zero-modes of the Dirac operator.

A major progress in understanding the manifestation of the index theorem on the
lattice was the realization that the eigenvectors of a $\gamma_5$-hermitian
lattice Dirac operator with real eigenvalues should be interpreted as the
lattice counterparts of the continuum zero-modes. This can be understood  from
the fact that only eigenvectors $\psi$ with real eigenvalues can have
non-vanishing pseudoscalar matrix elements $\langle\psi|\gamma_5|\psi\rangle$,
like the zero-modes in the continuum. 

Continuum QED in 4D in its usual non-compact realization for the gauge field 
has no topological charge.  For the lattice formulation with compact
representation of the gauge fields on the other hand the situation changes.
Lattice QED exhibits a twofold phase structure: a physical one, containing the
massless photon and a confining phase with properties similar to QCD. Some
phenomena appearing in the confinement phase are: zero modes, magnetic
monopole-antimonopole pairs and the occurrence of a non-zero chiral condensate.

Chiral symmetry breaking is a key feature of the theory of strong interactions,
QCD.  The order parameter is the chiral condensate $\langle \bar{\psi}\psi
\rangle$; it is related to the density of eigenvalues $\rho (\lambda)$ of the
Dirac operator near the origin via the Banks-Casher relation \cite{BaCa80}.  As
compact lattice QED shows a confining phase for certain values of the coupling
constant one also expects \cite{SaSe}  and indeed observes
\cite{SaSe} the appearance of a chiral condensate.

For a study of zero-modes and chiral symmetry breaking it is advisable to work
with chirally symmetric Dirac operators.  On the lattice chiral symmetry is
realizable in a locally violated form, expressed through the Ginsparg-Wilson
relation \cite{GiWi82} which in its original form  reads:
\begin{equation}
\gamma_5 \,\mathcal{D}+\mathcal{D}\,\gamma_5=2\,a\,\mathcal{D}\,\gamma_5\, R\,\mathcal{D}\;. \label{eq:gw2}
\end{equation}
where $\mathcal{D}$ denotes the massless Dirac operator, $a$ is the lattice spacing and
$R$ a local function of the gauge fields. In the continuum limit $\mathcal{D}$
anticommutes with $\gamma_5$ thus showing chiral symmetry. Dirac operators satisfying
the Ginsparg-Wilson relation preserve a lattice version of chiral symmetry \cite{Lu98}
without fermion doubling.  For a given local operator $R$ there may be many Dirac
operators. The eigenvalue spectrum of a Dirac operator for $R=\frac{1}{2}$ lies on a unit
circle around the center $1/a$ and complex eigenvalues come in  complex conjugate
pairs. Only for real $\lambda$ the expectation value of $\gamma_5$ between the
eigenstates $\bra{\psi_\lambda}\gamma_5 \ket{\psi_\lambda}$ is non-zero. Therefore
exact zero-modes have definite chirality.

Although there are several realizations of the Dirac operator satisfying the
Ginsparg-Wilson relation approximately, only the numerically costly overlap
operator \cite{NaNe} is an exact realization. Within QCD there have been
several studies on the density of near-zero-modes and the properties of exact
zero-modes and their relationship to topological excitations (see e.g.
\cite{GaGoLa01,Ga02a,HoDoDr02,Ga03}).

In compact lattice QED there have been studies with the overlap operator
\cite{BeHeMa01} demonstrating the existence of zero-modes in the confined
phase and relating the density of the near-zero-modes to the universal distributions
from Random Matrix Theory. Here we want to continue the study of such modes with an
emphasis on the chirality and the localization properties of exact zero-modes and
possible correlations to topological structures like monopoles.

\section{Formalism}
  
The Wilson gauge action reads 
\begin{equation} 
S[U]=\beta \sum_{x, \mu>\nu} (1-\cos \theta_{x,\mu\nu})\;, 
\end{equation}
where the gauge fields are represented by group elements
$U_{x,\mu}=exp(i\theta_{x,\mu})\in U(1)$ with $\theta_{x,\mu}\in (-\pi,\pi]$. The
plaquette angles are given by $\theta_{x,\mu\nu} =
\theta_{x,\mu}+\theta_{x+\hat{\mu},\nu}- \theta_{x+\hat{\nu},\mu}-\theta_{x,\nu}\in
(-4\pi,4\pi)$. This compact realization of the gauge fields and the action introduces
higher order self-interactions. The pure gauge theory has a confinement phase for small
$\beta$ and a Coulomb phase above a phase transition which for this action is close to
$\beta\approx 1$ and weakly 1st order \cite{CaCrTa,ArLiSc01}
(for a recent study of the phase transition suggesting a new
order parameter cf. Ref. \cite{VeFo04}). One observes an
abundance of monopoles in the confined phase. Monopoles in compact $U(1)$ are defined 
following Ref. \cite{DeTo80} from deficit angles of the plaquettes bordering 3-cubes,
corresponding to links on the dual lattices. Due to current conservation the links form
closed loops  (better: networks \cite{KeReWe94}) on the dual lattice.

The massless  overlap operator $D_{\textrm{\scriptsize ov}}$ may be written
\begin{equation}
\mathcal{D}_{\textrm{\scriptsize ov}}= 1+\gamma_5\,\epsilon(H) \;, \label{eq:of1}
\end{equation}
where $\epsilon$ denotes the operator $sign$-function and $H=\gamma_5 (s-H_{0})$ is some
Hermitian Dirac operator, constructed from an arbitrary Dirac operator. It is convenient
to use for $H_0$ the usual Wilson Dirac operator with negative mass term and $s$ is a
real parameter  which can be adjusted such as to  optimize the convergence in the
construction of $\epsilon(H)$. If $H_0$ is already an overlap operator then 
$\mathcal{D}_{\textrm{\scriptsize ov}}=H_0$.

For a $U(1)$ gauge theory on manifolds with torus topology there is no topological
charge in the sense of a non-vanishing Pontryagin index. So the Atiyah-Singer index
theorem may be realized only trivially, i.e. allowing for a cancellation between the
numbers of left-handed and right-handed zero-modes. However, on the lattice we have
non-differentiable fields and therefore we may expect other violations as well.

It has been demonstrated in QCD that for instanton configurations the Dirac operator
shows a zero-mode localized in space-time  \cite{GaGoLa01,GaGoRa01,GaLa01,Ga02a}. To
our knowledge nothing is known about the localization properties  of zero-modes in
QED. 

In order to study the localization properties a simple set of gauge invariant
quantities  has been introduced. For an eigenvector $\psi(x)$ of the lattice Dirac
operator one defines a local density
\begin{equation}
p_\sigma(x)=\sum_{d}\psi (x)^*\Gamma_\sigma\psi (x) \;,
\label{eq:loc1}
\end{equation}
where $\Gamma_\sigma$ is an element of the Clifford algebra and the local
sum runs  over the Dirac indices. The
eigenvectors are normalized to unit norm, i.e. 
\begin{equation}
\sum_{x} p_0(x) =\langle\psi^{\dag} \psi \rangle=1\;. \label{eq:loc3}
\end{equation}
For the  cases of particular interest we abbreviate the scalar density $p(x)=p_0(x)$
(for $\Gamma_0$ the unit matrix) and the chiral density $p_5(x)$  (for $\gamma_5$). We
also determined the vector densities $p_\mu$ (for $\Gamma=\gamma_\mu$), the axial
vector densities $p_{5\,\mu}$  (for $\Gamma=\gamma_5\gamma_\mu$) and the  tensor
densities $p_{\mu\nu}$ (for $\Gamma=\gamma_\mu\gamma_{\nu\neq\mu}$).

Exact zero-modes are also eigenmodes of $\gamma_5$ with eigenvalues $\pm 1$.
Therefore for such eigenmodes we expect even locally
\begin{equation} \label{pp5relation}
p(x)=\pm p_5(x)\;.
\end{equation}
The integral over the chiral density provides a measure for the amount of chiral
symmetry breaking because it takes on its largest value for the zero-modes, namely $\pm
1$ for normalized eigenvectors, and vanishes otherwise. 

The so-called \emph{inverse participation ratio} (IPR) is introduced for further
quantification of the localization (see for example Ref. \cite{GaGoLa01}). We define it for
both, $p$ and $p_5$:
\begin{equation}\label{eq:loc4}
I=V\sum_{x}\;p(x)^2\;,\;\;
I_5=V\sum_{x}\;p_5(x)^2,
\end{equation}
where $V$ is the lattice volume. 

Assume that the density is evenly distributed on a subvolume $V_f$  with $p(x)=1/V_f$
and vanishing elsewhere. Then we find $I=V/V_f$,  with the limiting cases $I=1$ for
distribution over the whole lattice and $I=V$ for localization at one point. It is
essentially the inverse fraction of  the volume contributing to the mode. This makes it
an appropriate measure for the localization of eigenmodes. 

Let us try to understand why the inverse participation ratio $I_5$ may be considerably
smaller than $I$ for near-zero-modes than for exact zero-modes. The local Dirac sum may
be written as a sum of two positive terms $p(x)=p_+(x)+p_-(x)$ where
$p_5(x)=p_+(x)-p_-(x)$. Therefore  $p(x)\ge |p_5(x)|$. Due to this property $I_5$ is
bounded for all eigenmodes: $I \ge I_5$. For exact eigenmodes we have $I=I_5$ due to
(\ref{pp5relation}).

Consider now eigenmodes corresponding to eigenvalues far from the origin: $p_5(x)$ will
be much smaller than $p(x)$ and fluctuate around zero. In this case $I_5$ is expected
to be significantly smaller than $I$. When the eigenvalues approach zero $I_5$
increases and the ratio $I_5/I$ is expected to approach $1$.

We also determine different densities $p_\sigma$ for exact zero-modes of
chirality +1 or -1. This amounts to checking relations like  e.g. 
\begin{equation}
\braket{\bar{\psi}}{\gamma_\mu\gamma_5\psi}=\pm
\braket{\bar{\psi}}{\gamma_\mu\psi}\;,
\end{equation}
which we could verify.
 
\section{Simulation and results}

The background gauge fields were obtained by using the Metropolis and overrelaxation
updating algorithm. All configurations are well decorrelated, separated by $5000$
updating  sweeps. We analyzed $400$ configurations on the $4^4$, $6^4$ and $8^4$
lattices and $100$ on  $12^4$ lattices. For the computation of the eigenvalues and
eigenvectors the so-called  implicitly restarted Arnoldi method \cite{SoLeSoYa}
was used.  The overlap operator was computed by an appropriate series expansion. Only
the smallest $10$ eigenvalues and their eigenvectors were calculated. Real eigenmodes
at zero have exact values $\braket{\bar{\psi}}{\gamma_5\psi} = \pm 1$.

\begin{figure}[t]
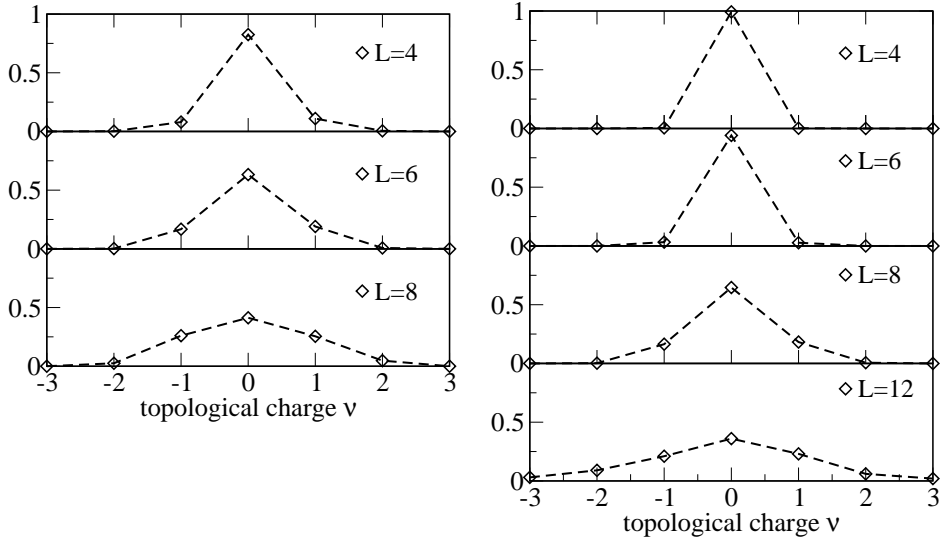

\begin{center}
\begin{tabular}{cc}
\parbox[b][7.1cm][t]{6cm}{\includegraphics*[width=6cm]{09coll.eps}}&
\includegraphics*[width=6cm]{099coll.eps}
\end{tabular}
\end{center}
\caption{The plots show the normalized histogram 
(probability distribution) for the topological charge
$\nu$  for the configurations at 
$\beta=0.9$ (left-hand plot) and $\beta=0.99$ (right-hand plot). 
\label{fig-zmd}}
\end{figure}

In the confined phase many exact zero-modes were observed whereas no zero-modes 
appeared in the Coulomb phase. Zero modes are chiral eigenmodes and are related to the
topological charge via the Atiyah-Singer index theorem. In our discussion we therefore
identify the number of zero-modes with positive or negative chirality in a
configuration with the positive or negative topological charge $\nu$. We never observed
configurations which have zero-modes of different chirality.  To our knowledge there is
no formal theorem explaining this feature in 4D (there exists such a vanishing theorem
in 2D \cite{AnKiNiSc}).

In Fig.\ \ref{fig-zmd} we show the number $\nu$ of  exact zero-modes with   chirality
$\pm 1$ for two different values of $\beta$. We observe obvious volume dependence. The
number of zero modes as well as the average percentage of zero-mode configurations
compared to all configurations grows with the volume. The integrated number of
zero-modes (i.e. adding up all $|\nu|$) $n_{zeros}/n_{configs}$ reaches values
$\mathcal{O}(0.9)$ for the large lattices. This can be observed for both $\beta=0.99$
and $\beta=0.9$.  On $4^4$ lattices at $\beta=0.99$  we found only 3 such modes on all
400 configurations (compared to e.g. 24 zero-modes on $6^4$). This small number can
possibly be attributed to the situation that the pseudocritical point (i.e. the peak
position of the specific heat and other cumulants)  on smaller lattices effectively
occurs already at smaller $\beta$-values and thus we may have been already in or at
least closer to the Coulomb phase for that lattice size. For $6^4$ the position of the
pseudocritical point is near 1.002 whereas for $12^4$ it is near 1.010, cf. Ref.
\cite{ArLiSc01}.

\begin{table}
\begin{tabular}{rrrrrrl}
\hline
$l$ & 
$\#_{conf}$ & 
$n(\nu=0)$ & 
$n(\nu=1)$ & 
$n(\nu=2)$ & 
$n(\nu=3)$ &
$10^4\times\langle \nu^2\rangle/V$\\
\hline
    4 & 400  &397  &  3   & 0   & 0  &\quad 0.29(17)\\
    6 & 400  &376  & 24   & 0   & 0  &\quad 0.46(9)\\
    8 & 500  &323  &173   & 4   & 0  &\quad 0.92(6)\\
   12 & 100  & 36  & 44   &15   & 5  &\quad 0.72(10)\\
\hline
\end{tabular}
\caption{\label{tab:susc}
Summary of configuration number with topological charge for $\beta=0.99$
and the corresponding topological susceptibility.}
\end{table}

Table\ \ref{tab:susc} summarizes the results for $\beta=0.99$ together with the
topological susceptibility. Its behavior is compatible with 
approaching a constant for large volumes.

Following the definitions from Refs. \cite{DeTo80,GrJaJe85} we determined the monopoles
(which are closed loops on the dual lattices). We find no correlation bet\-ween the
monopole density and the number of exact zero-modes on the corresponding
configurations. For  $\beta=0.9$ the monopole density is 0.19(1) for both lattices
sizes $6^4$ and $8^4$, for $\beta=0.99$  it varies from 0.118 to 0.121 for $|\nu|$
ranging from 0 to 3 on the largest lattice size $12^4$. For $\beta>\beta_c$ the density
decreases exponentially. We see no correlation between the number of zero-modes  and
the integrated  monopole loop lengths at $\beta=0.9$ and 0.99; this observation
confirms \cite{BeHeMa01}.

The scaling of the density of small near-zero-modes in the sector of different zero
mode number has been compared with Random Matrix Theory expectations in Ref.
\cite{BeHeMa01}. The universality class was identified to be  the unitary
ensemble. We also studied the  densities for our data and agree with these findings.
Due to our choice of a different offset parameter value $s$ in the Wilson action
entering the overlap operator construction, the condensate value (and with it the
scaling parameter) has to be renormalized multiplicatively \cite{KiNaNe98} for the 
comparison.

\begin{figure}[t]
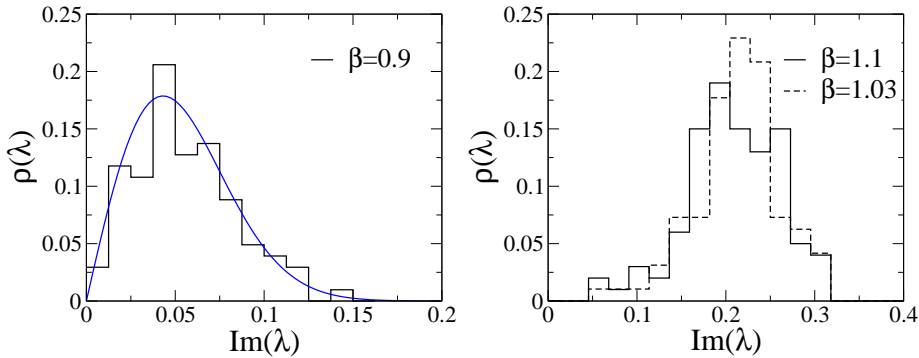

\begin{center}
\begin{tabular}{cc}
\includegraphics*[width=6cm,clip=true]{rho_b090_L8_finest_pluscurve.eps}
\includegraphics*[width=6cm]{rho_b103_b110_L8.eps}
\end{tabular}
\end{center}
\caption{We compare the densities of the smallest eigenvector for $\beta=0.9$ 
(in the $\nu=0$ sector) and $\beta=1.03$ and $1.1$ for lattices size $8^4$.
\label{fig:density}}
\end{figure}

Fig. \ref{fig:density} compares the densities for the confinement region with
those for the Coulomb region, exhibiting a different behavior. In the
confinement region one expects for $\nu=0$ a distribution following the chiral
unitary ensemble of Random Matrix Theory
\cite{NiDaWe98} and we overlay a fit to that functional behavior
\begin{equation}
\rho(z)=\frac{z}{2}\,\exp{(-z^2/4)}\;.
\end{equation}
In the Coulomb region the results for $\beta=1.03$ and $\beta=1.1$ are compatible with
each other indicating that there is no gap, different to the observed behavior in the
deconfined phase in QCD. Our statistics is not sufficient to exclude a small gap,
though.

For the zero-modes we observe $p(x)=\pm p_5(x)$ as expected for exact GW-fermions and
thus also the IPRs agree. For chiral eigenmodes we expect local agreement of those
densities which correspond to relations for the gamma matrices like
$\gamma_1\,\gamma_2=-\gamma_3\,\gamma_4\,\gamma_5$, i.e. $p_{12}(x)=-p_{34}(x)$. Indeed
the tensor densities for the chiral modes do obey this relation within their group. The
local densities for vector and axial vector both vanish. For non-zero-modes these
relations are not valid.

In general the integrated tensor densities are comparatively small in absolute
magnitude ($0.1\ldots 0.01$) for zero-modes. For near-zero-modes all integrated vector,
and tensor densities are small ($0.1\ldots 0.01$). The integrated  pseudoscalar density
was exactly $\pm 1$ as expected for exact  zero-modes. 
 
\begin{figure}[t]
\begin{center}
\includegraphics*[height=6cm]{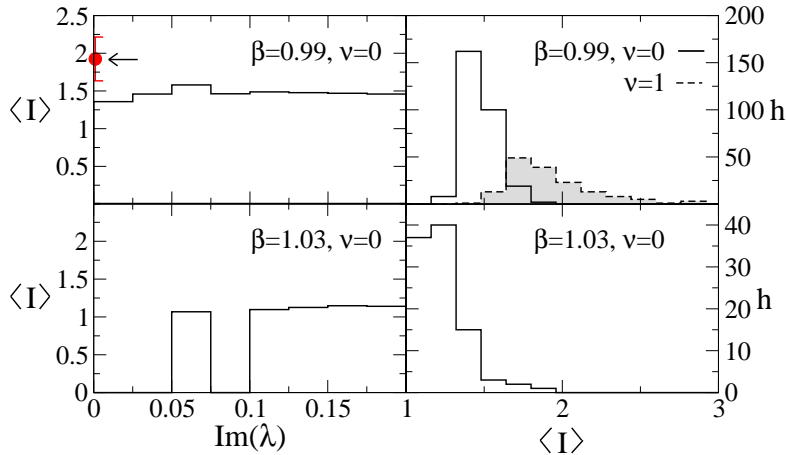}
\end{center}
\caption{\small The graphs show the mean inverse participation ratio for the
near-zero-modes (in the $\nu=0$ sector) at $\beta=0.99$ and $\beta=1.03$ (lattice size
$8^4$): mean values of IPR (l.h.s.) and IPR-histograms (r.h.s.).  The exact zero modes
in the $\nu=1$ sector have a wider distribution (shaded area) extending to higher
values; the $\langle I\rangle$ and s.d. for those is indicated by the filled circle
(arrow) and error bars in the upper l.h.s. plot. \label{fig:ipr}}
\end{figure}

The inverse participation ratio should allow some information on the space-time
localization of the eigenmodes. In Fig.\ \ref{fig:ipr}  we summarize this
quantity for the modes for {\em non-zero} eigenvalues, binned according to
Im($\lambda$) and also show the IPR histograms. In the confinement region we
find no increase of the localization for the lowest bins. This observation
differs from results in QCD studies \cite{GaGoRa01}, where also the near-zero
modes show stronger localization.

In the Coulomb region no modes at very small $\textrm{Im}(\lambda)$ were observed and
the low-lying modes show small IPR, close to the minimum of 1.

Let us now concentrate on the {\em exact} zero-modes in the confining phase. These are
well localized objects with comparatively large individual IPR values  up to 4.2.  For
fixed $\beta$  the $\langle I\rangle$ show little if any  volume dependence with values
of 1.92(29) for $L=8$ and 2.03(69) for $L=12$ (this independence on the volume was also
observed in recent QCD studies \cite{AuBeGo04}). These values are, however, clearly
larger than the mean values for the small non-zero modes. The difference is mainly due
to an IPR distribution extending to much higher individual IPR values, as seen in Fig.
\ref{fig:ipr}. The values of IPR for the lowest non-zero modes for the sector with no
or that with one exact zero mode agree within small errors.

\begin{figure}[tp]
\begin{center}
$\begin{array}{c@{\hspace{0.1cm}}c@{\hspace{0.1cm}}c@{\hspace{0.1cm}}c}
        \epsfig{file=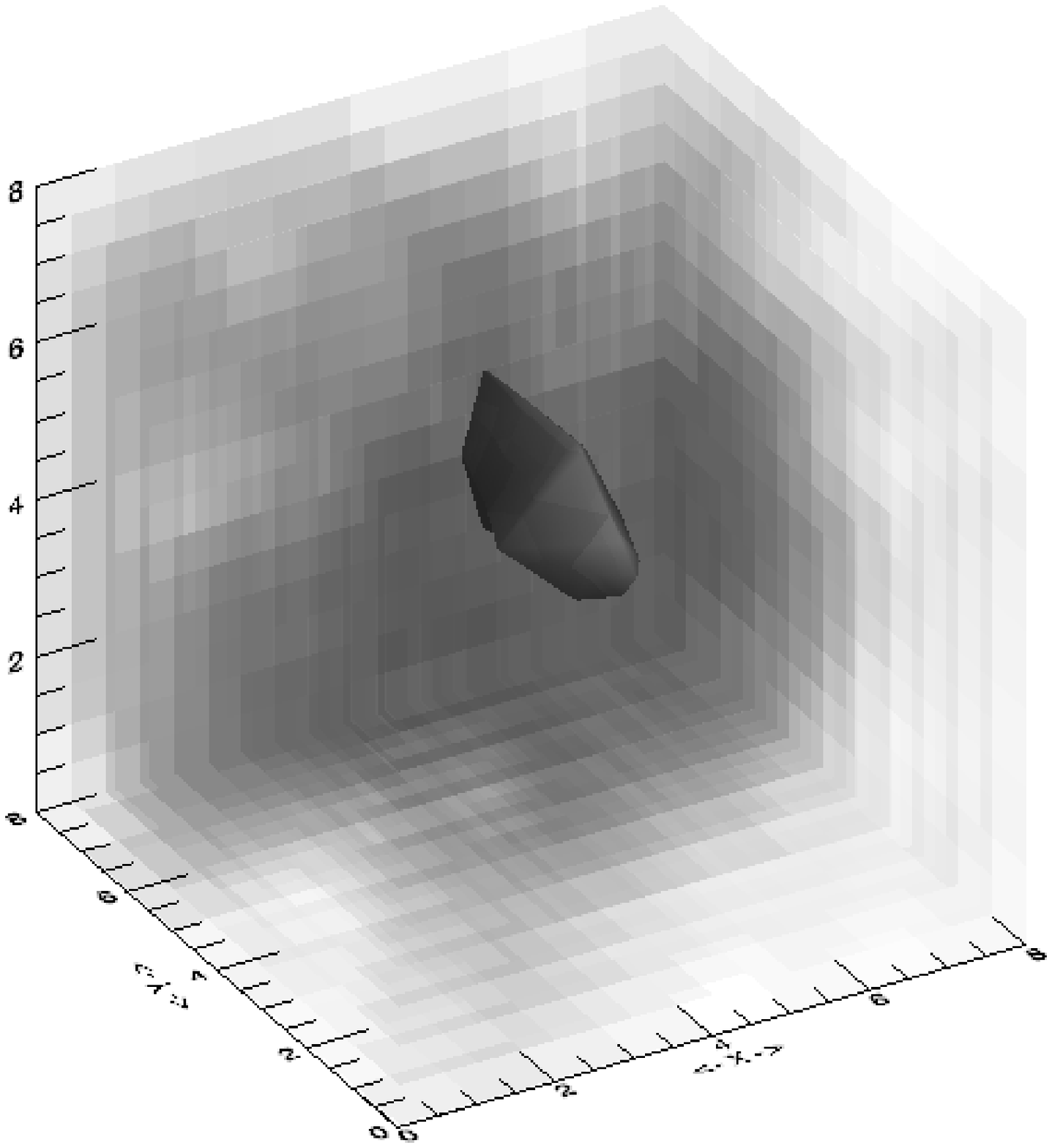, 
	     bb=40 70 520 610, clip=, scale=0.12} &
	\epsfig{file=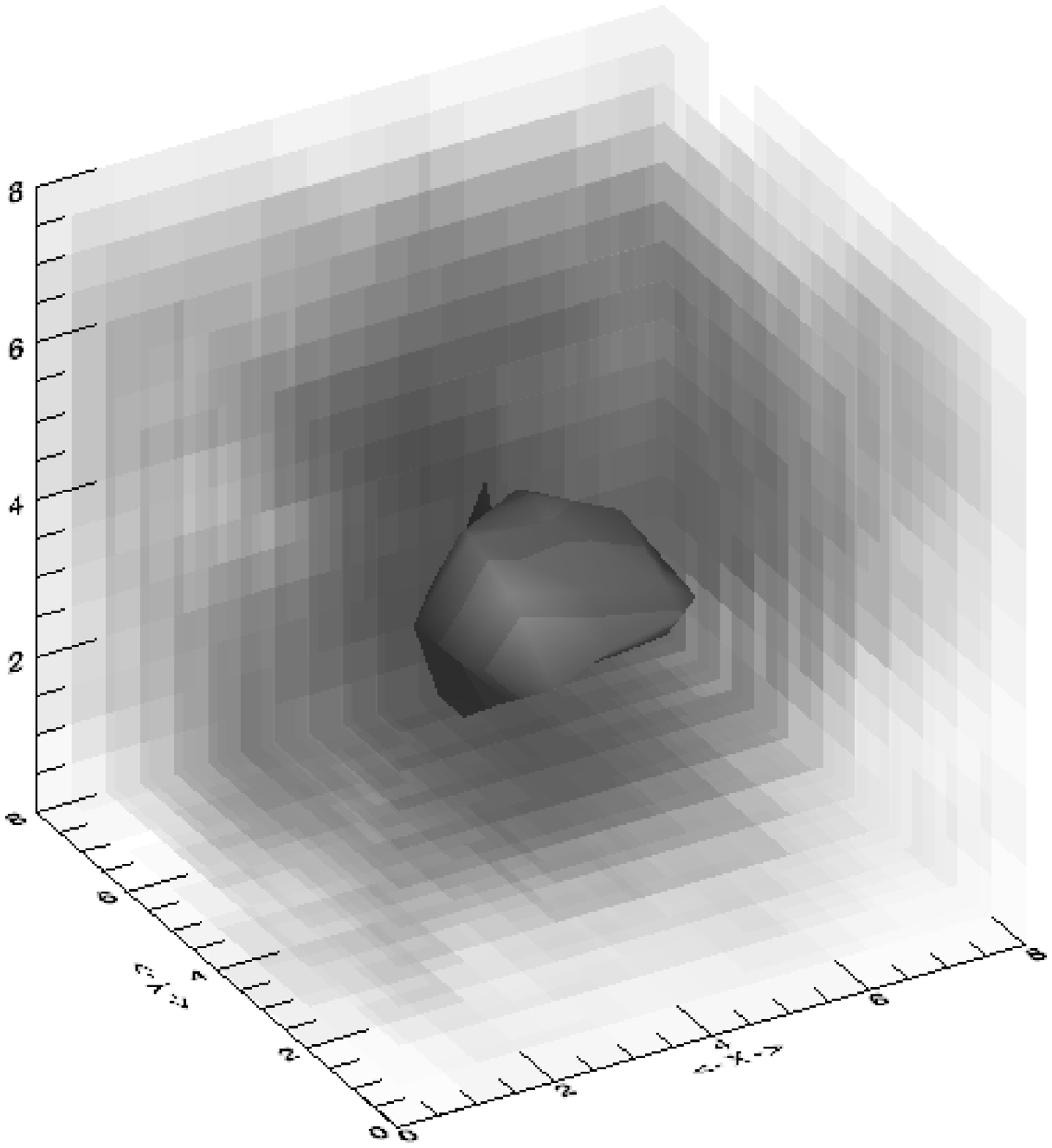, 
	     bb=40 70 520 610, clip=, scale=0.12} &
	\epsfig{file=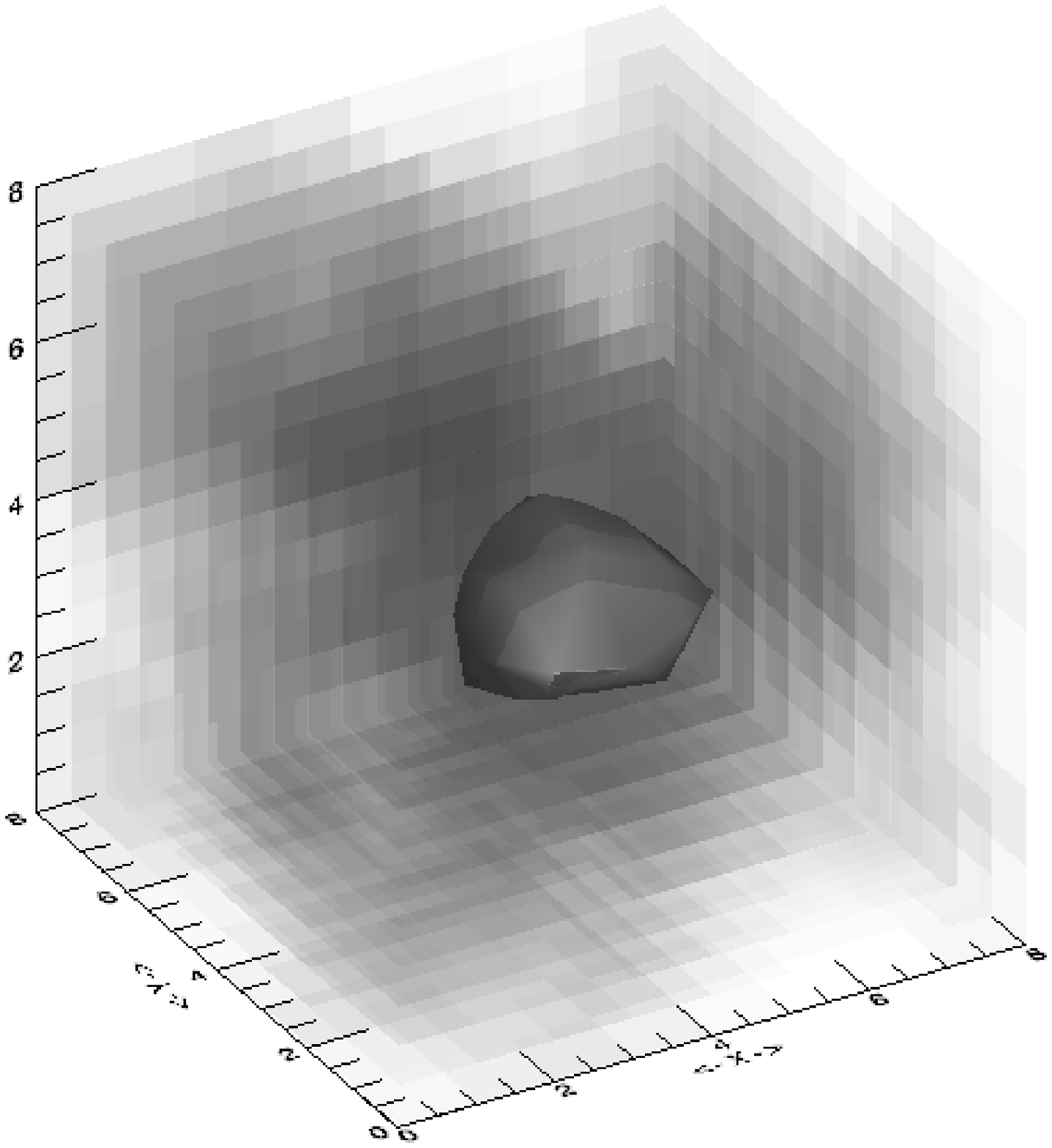, 
	     bb=40 70 520 610, clip=, scale=0.12} &
	\epsfig{file=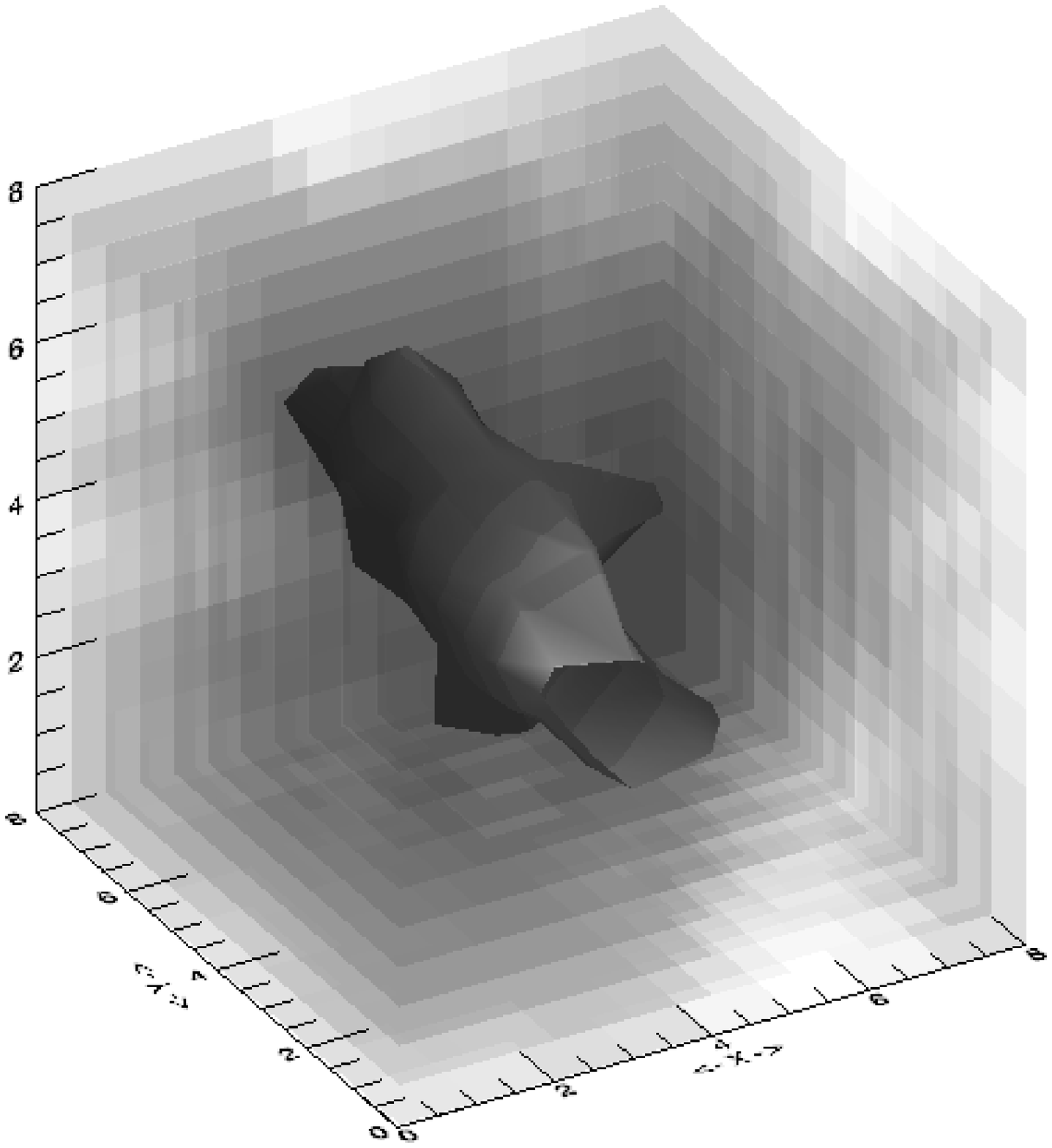, 
	     bb=40 70 520 610, clip=, scale=0.12} \\ 
	\epsfig{file=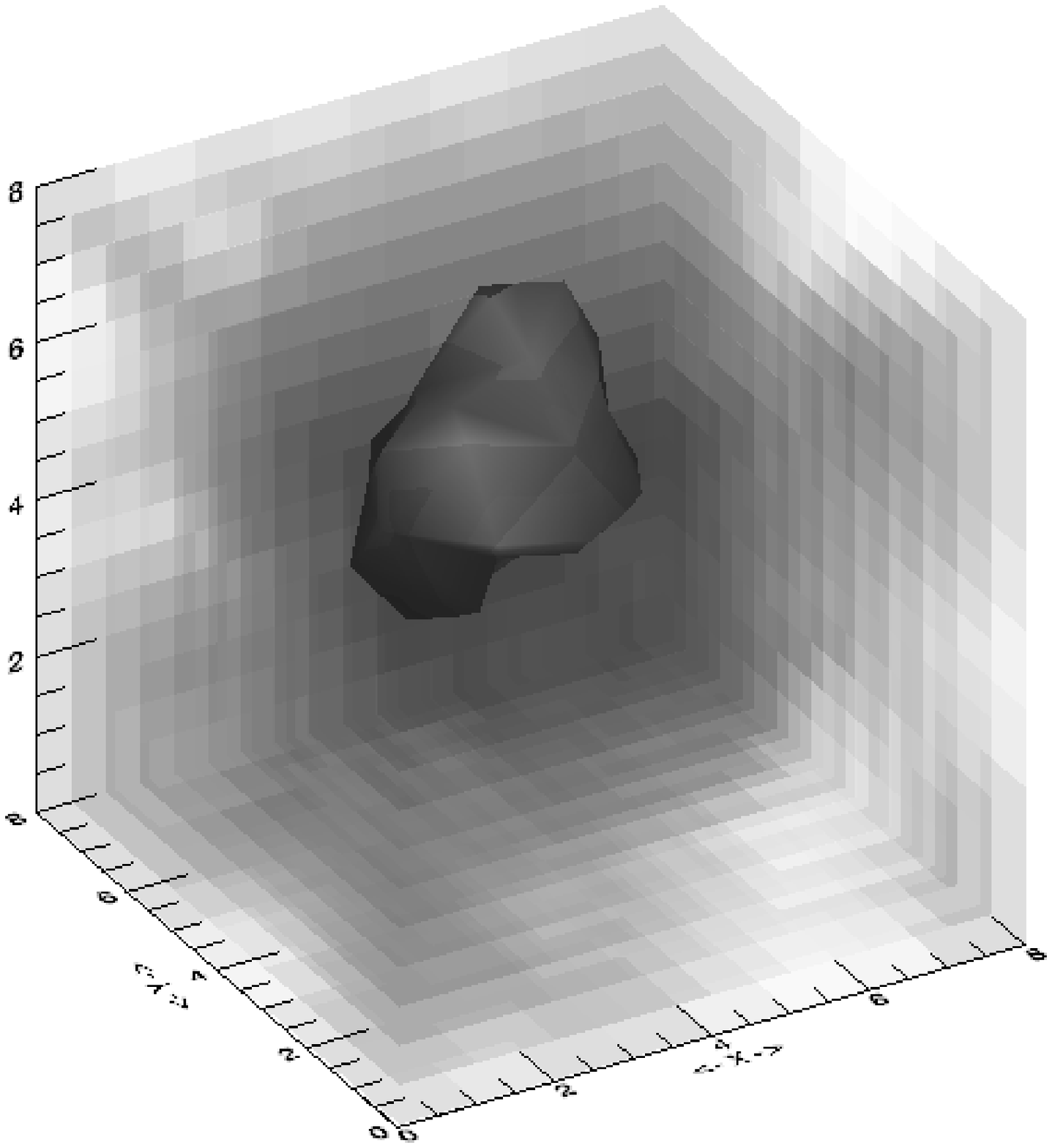, 
	     bb=40 70 520 610, clip=, scale=0.12} &
	\epsfig{file=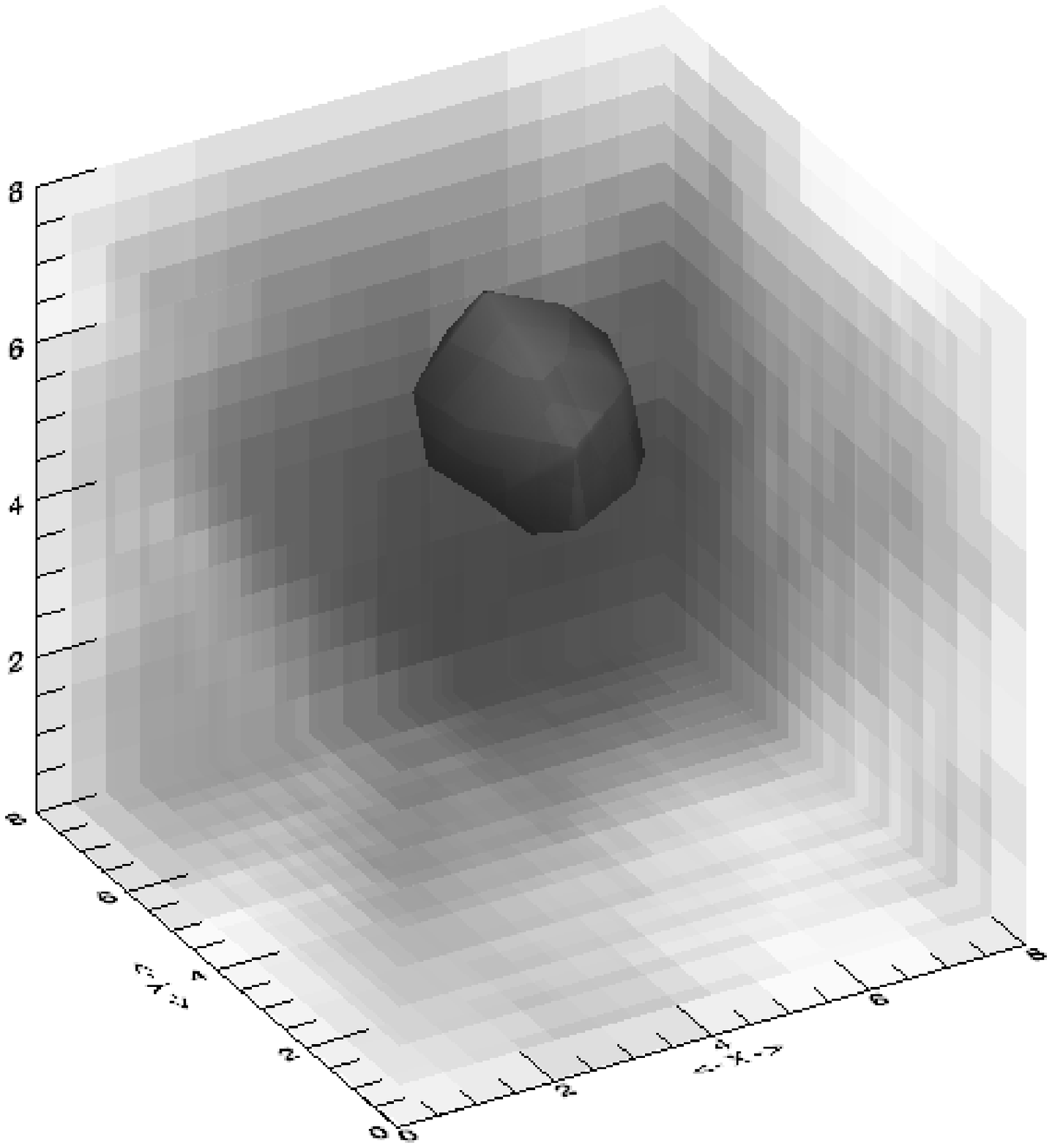, 
	     bb=40 70 520 610, clip=, scale=0.12} &
	\epsfig{file=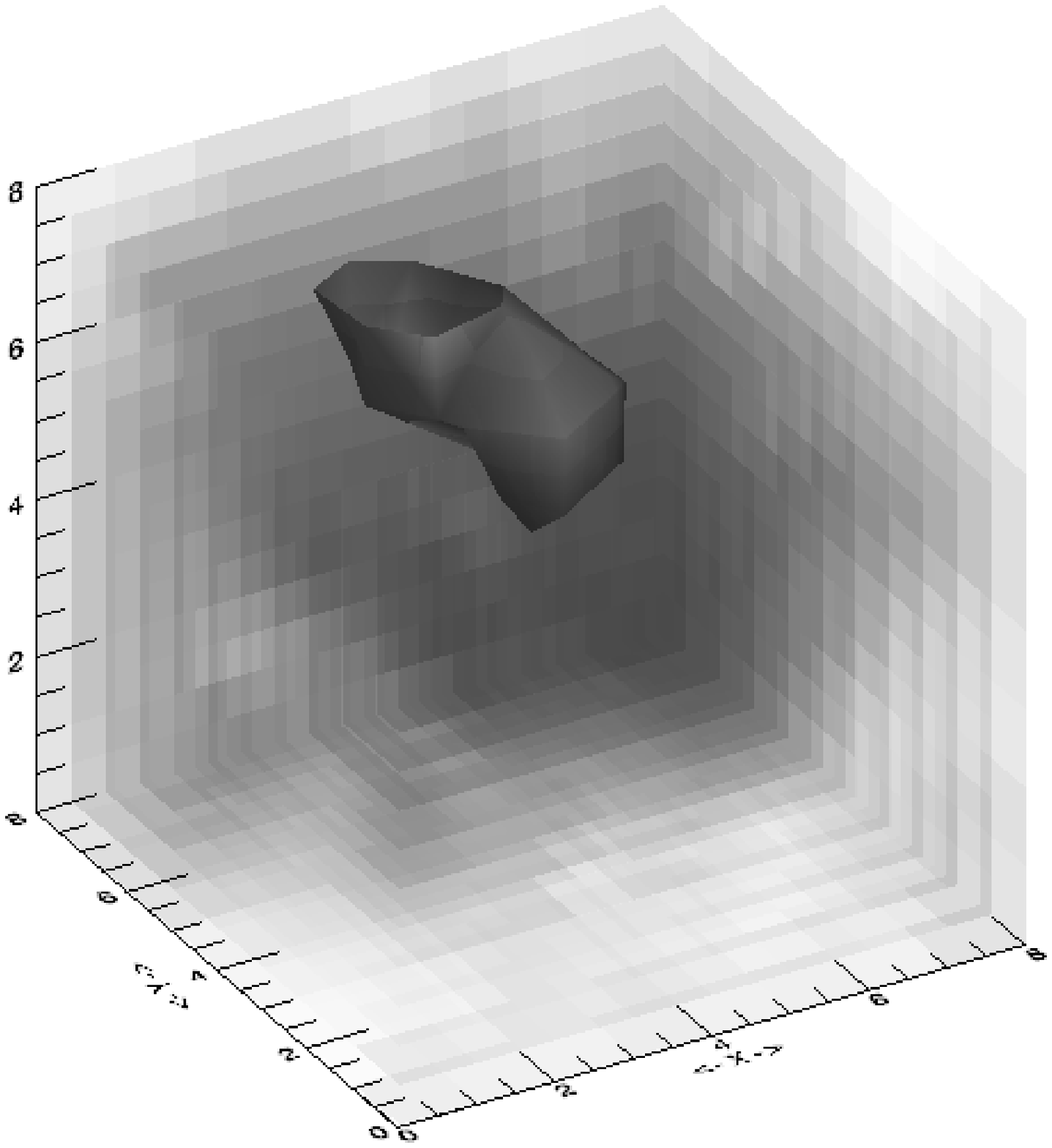, 
	     bb=40 70 520 610, clip=, scale=0.12} &
	\epsfig{file=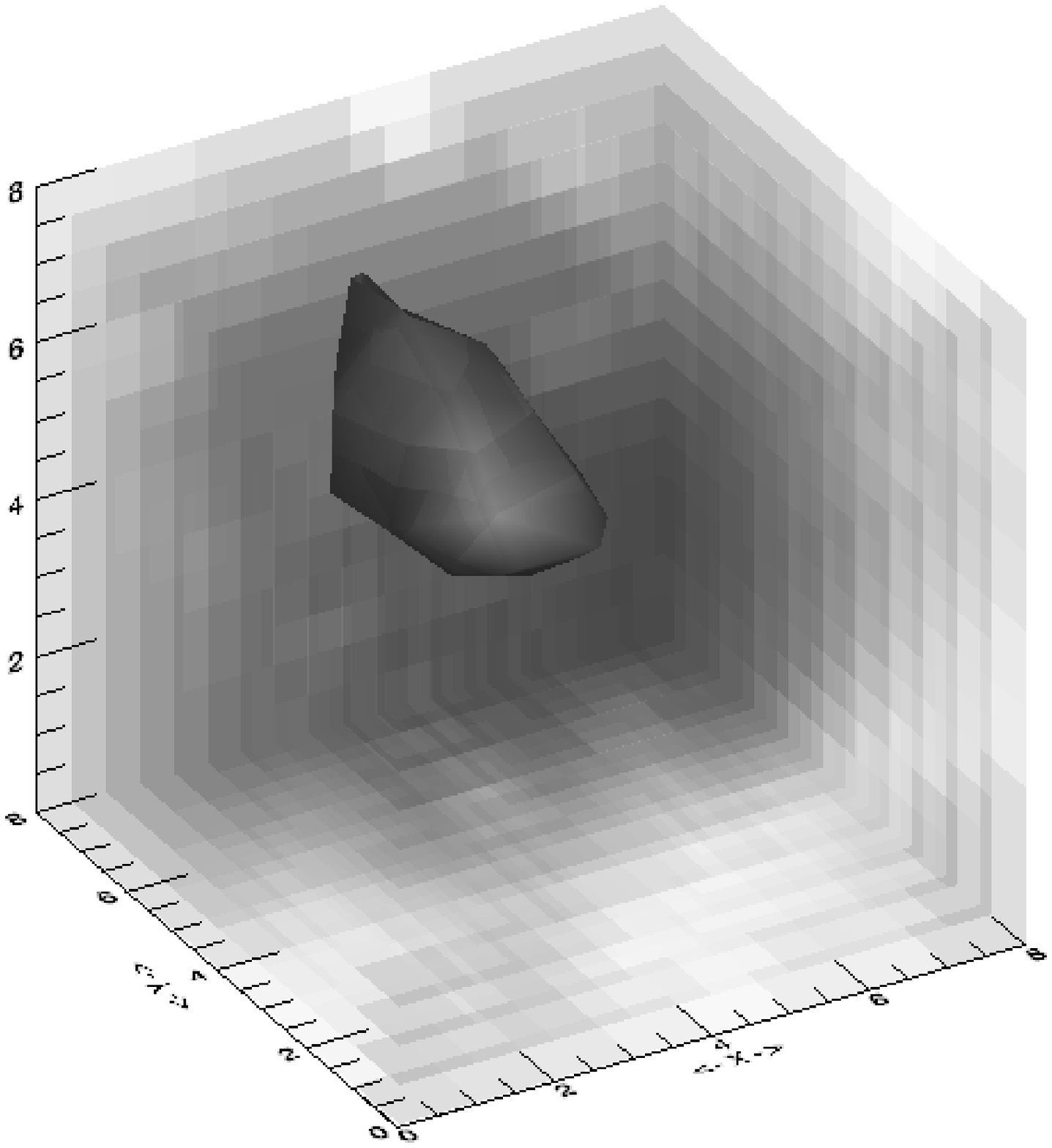, 
	     bb=40 70 520 610, clip=, scale=0.12} \\                                 
\end{array}$
\end{center}          
\caption{\label{fig:MCdensity}The density shape for zero-mode eigenvector of a Monte
Carlo generated gauge configuration (at $\beta=0.99$, lattice size $8^4$).  The 3D cuts
are for time slices $n_t=1\ldots 4$ (first row) and $n_t=5\ldots 8$ (second row). This
zero-modes has $I=2.75$.}
\end{figure}

In an attempt to visualize the geometrical shape of these localized zero-modes we
studied 3D cuts of the density distribution. Fig.\ \ref{fig:MCdensity} gives an example
of the observed structure.  We plot the surfaces of constant density $p(x)$ for an
exact zero-mode for the eight 3D time slices ($n_t=1\ldots 8$) of the $8^4$ lattice.
The shape describes a tubular structure in 4D. However, most of the observed shapes are
less clear and often made up from disconnected pieces. Contrary to QCD \cite{GaLa01} no
clear instanton-like picture (i.e. 4D ``blobs'') evolves. The structure we observe for
the zero-mode eigenvectors are sometimes ball-like in 4D, sometimes tube-like, even
tubes closed in some  direction due to the periodic boundary conditions. Although a
relation to the concept of closed monopole loops (cf. Ref. \cite{GrJaJe85}) may be
conjectured, no stringent conclusion can be drawn.

This evidence supports the notion, that the zero-modes are not instanton-like lumps,
but extended and periodically closed in one or more directions.  In a recent study 
\cite{AuBeGo04}  of the zero-modes in QCD it was argued, that the scaling of  $\langle
I\rangle$ with the lattice spacing may be used to identify the co-dimension of these
objects. Due to lack of results at several different lattice spacing we cannot attempt
this here. In  Ref. \cite{AuBeGo04} values of $\langle I\rangle\approx 2.8$ were observed,
i.e. larger than our values $\langle I\rangle\approx 2$, indicating even stronger
localization.

In order to gain further insight, we determined for some gauge configurations the
eigenmodes of the Dirac operator applying also anti-periodic boundary conditions in the
time direction by multiplying all temporal gauge links of a timeslice with -1. This
leaves the gauge action invariant (although it is not a gauge transformation on the
finite lattice).  The number of monopoles and Dirac plaquettes does not change,
either. 

However, the  characteristic polynomial coefficients of the eigenvalue equation for the
Dirac matrix involve traces over closed loops and thus may differ due to the
periodically closed loops.  The Dirac matrix was diagonalized again for the such
modified gauge configuration and the eigenvalues were compared with the ones without an
added phase: Quite often the zero-modes disappeared. Where the zero-mode survived the
transformation, the $3$-D density of the  new eigenvectors had a completely different 
structure. This confirms the suspicion that for $U(1)$ gauge theory the zero modes are
related to lumps extending in at least one direction.  Changing the boundary conditions
in QCD sometimes also repositions topological objects 
(see e.g. \cite{GaPuGaSo} and references therein) and in a recent analytical study it was
reported \cite{Ga04} that zero modes with constant curvature may be constructed in the
$U(1)$ subgroup. These modes may be ``switched'' on and off by changing the boundary
conditions. This is in agreement with our observation.

\begin{figure}[tp]
\begin{center}
$\begin{array}{c@{\hspace{0.1cm}}c@{\hspace{0.1cm}}c@{\hspace{0.1cm}}c}
	\epsfig{file=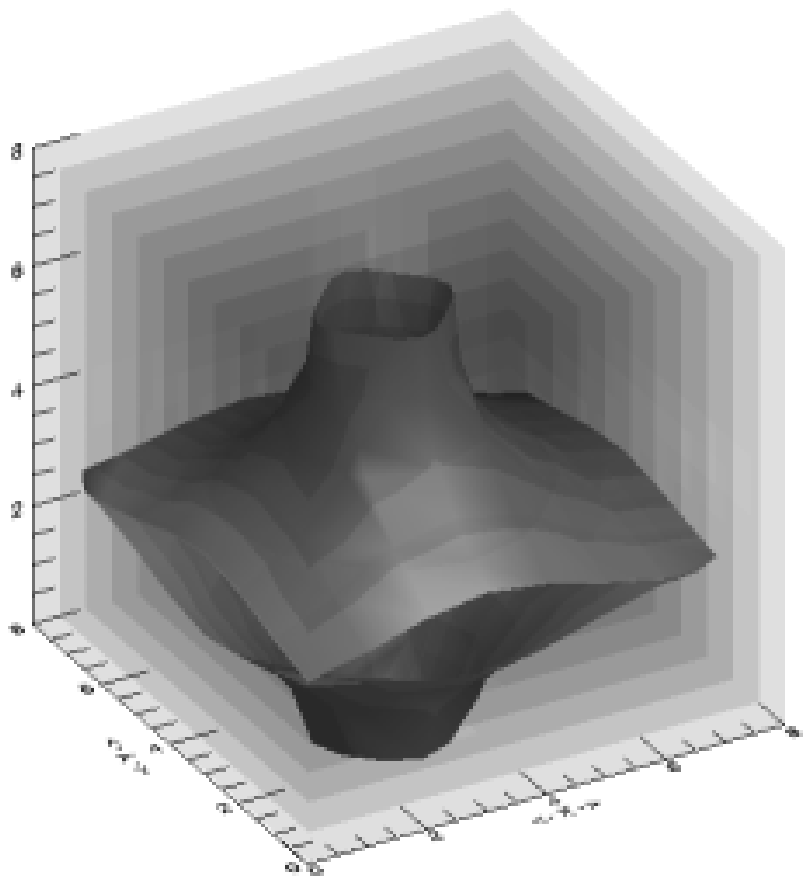,
	     bb=15 30 290 330, clip=, scale=0.25} &
	\epsfig{file=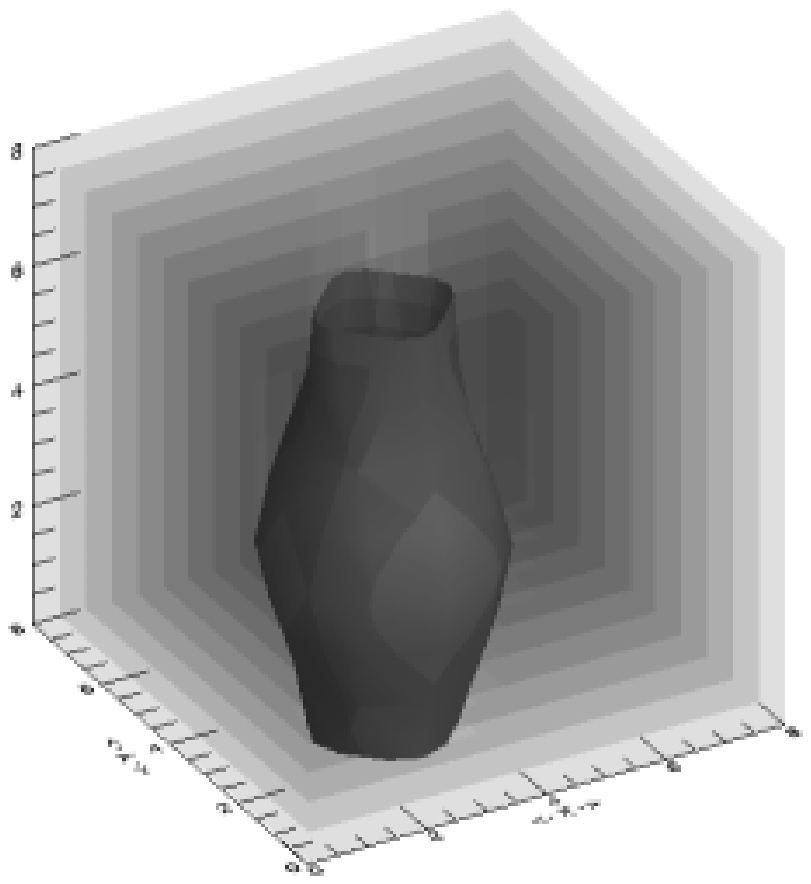,
	     bb=15 30 290 330, clip=, scale=0.25} &
	\epsfig{file=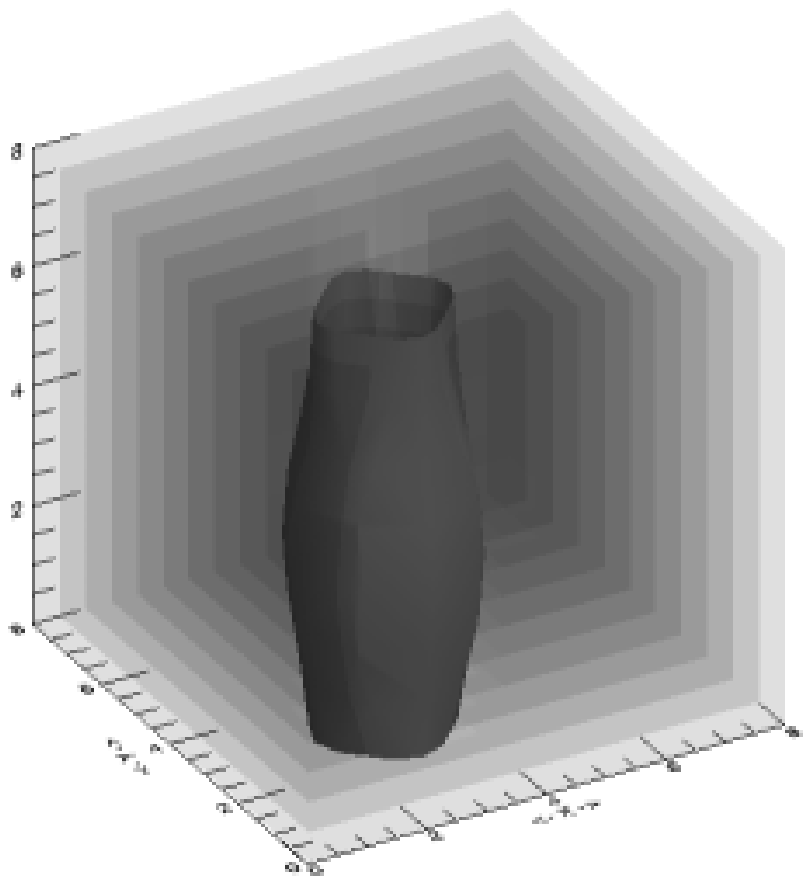,
	     bb=15 30 290 330, clip=, scale=0.25} &
	\epsfig{file=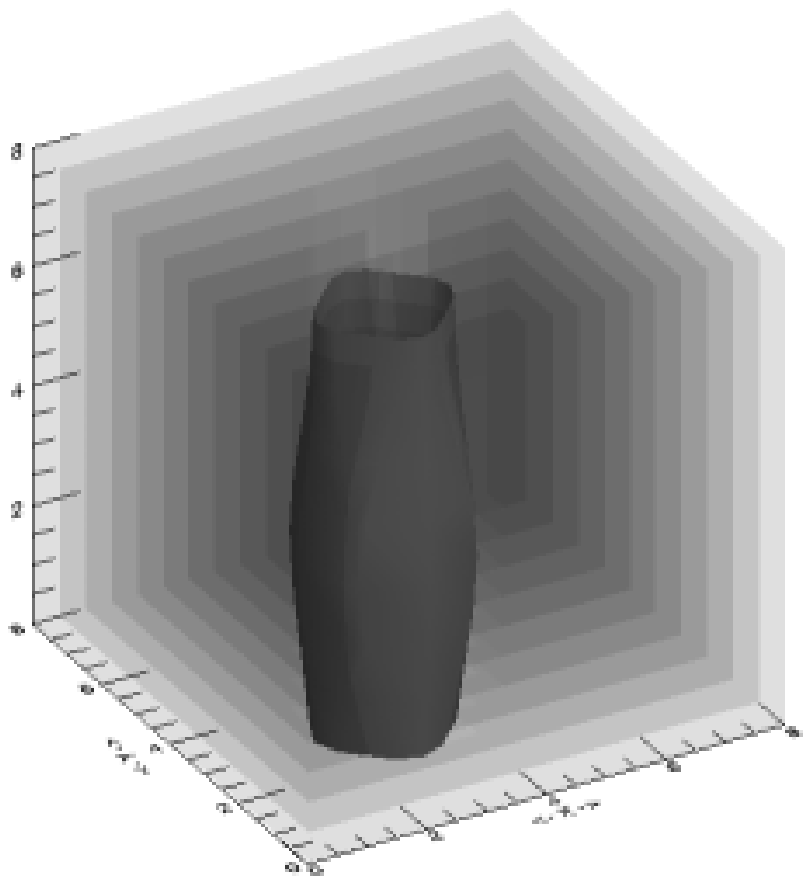,
	     bb=15 30 290 330, clip=, scale=0.25} \vspace{-3mm} \\ 
	\epsfig{file=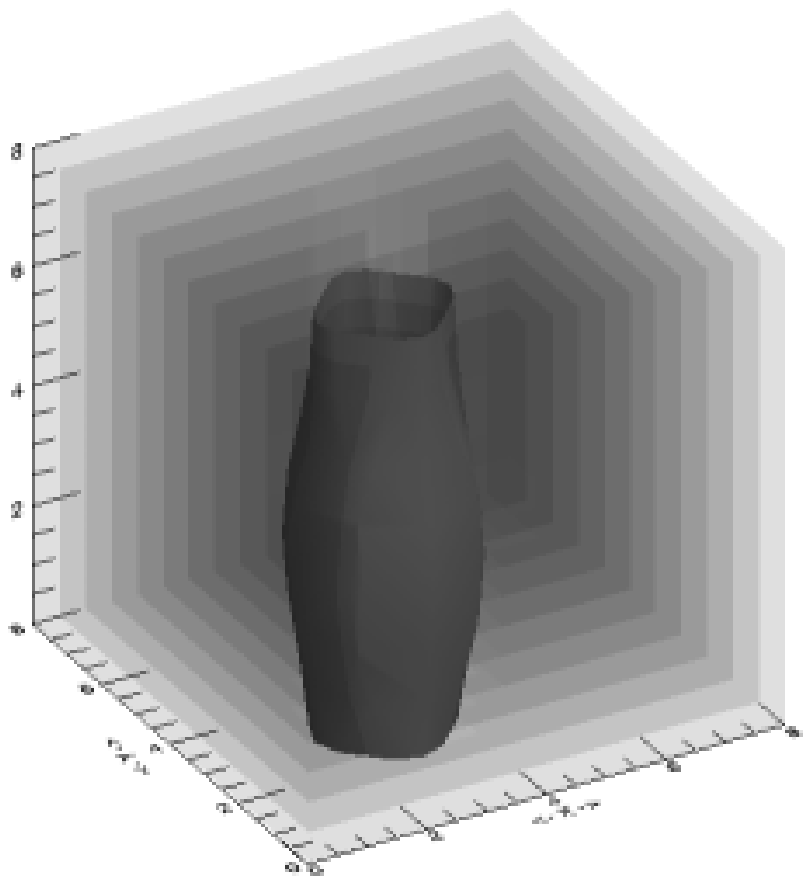,
	     bb=15 30 290 330, clip=, scale=0.25}  &
	\epsfig{file=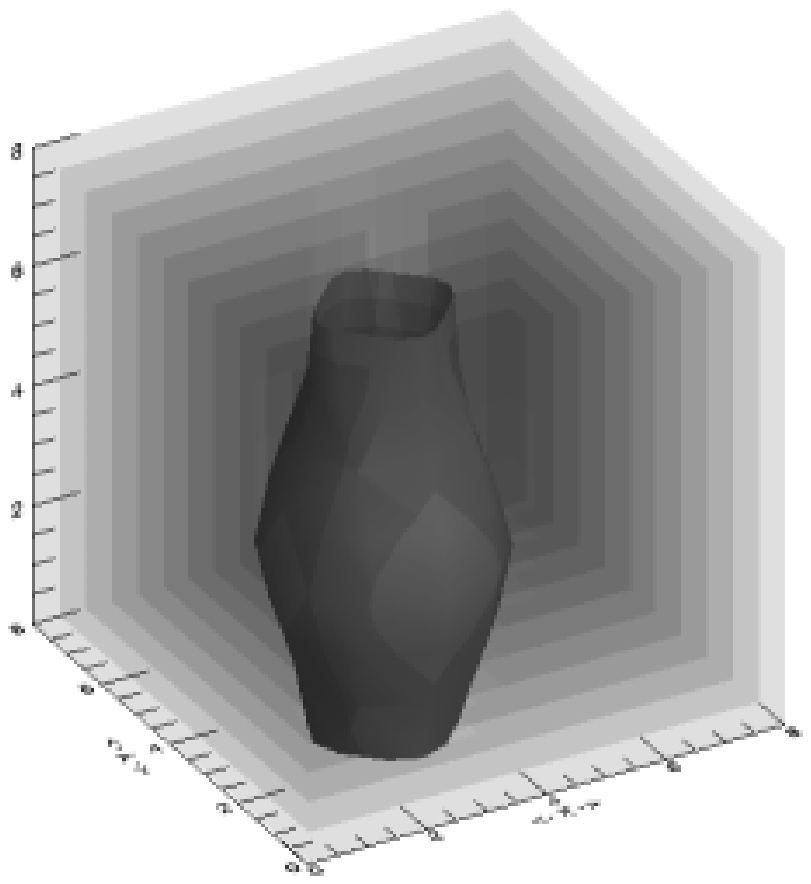,
	     bb=15 30 290 330, clip=, scale=0.25}  &
	\epsfig{file=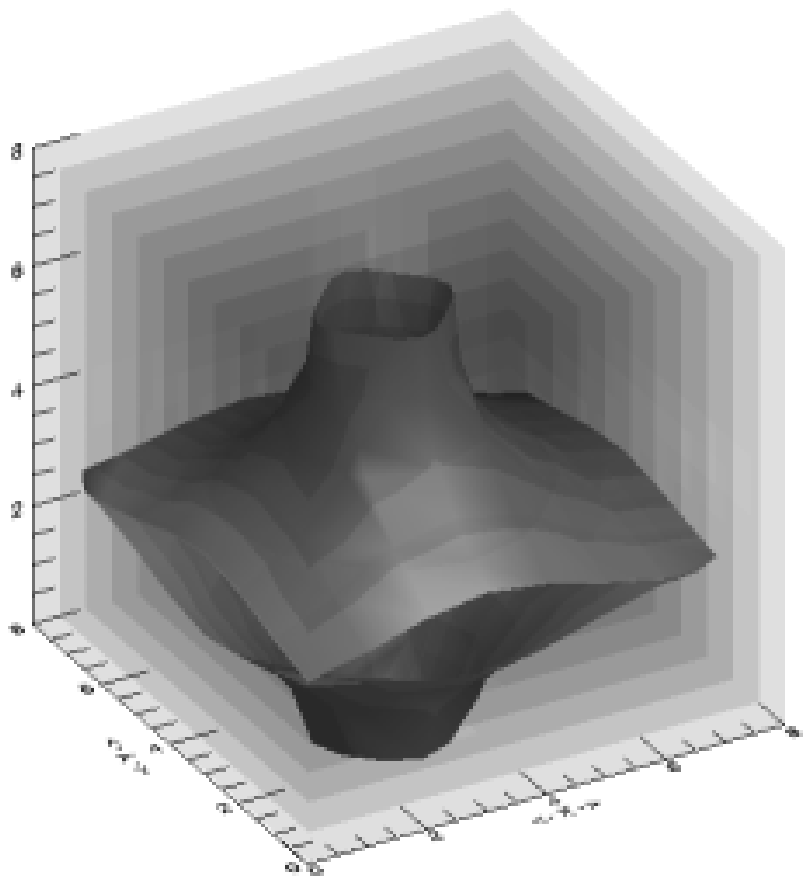,
	     bb=15 30 290 330, clip=, scale=0.25}  &
	\epsfig{file=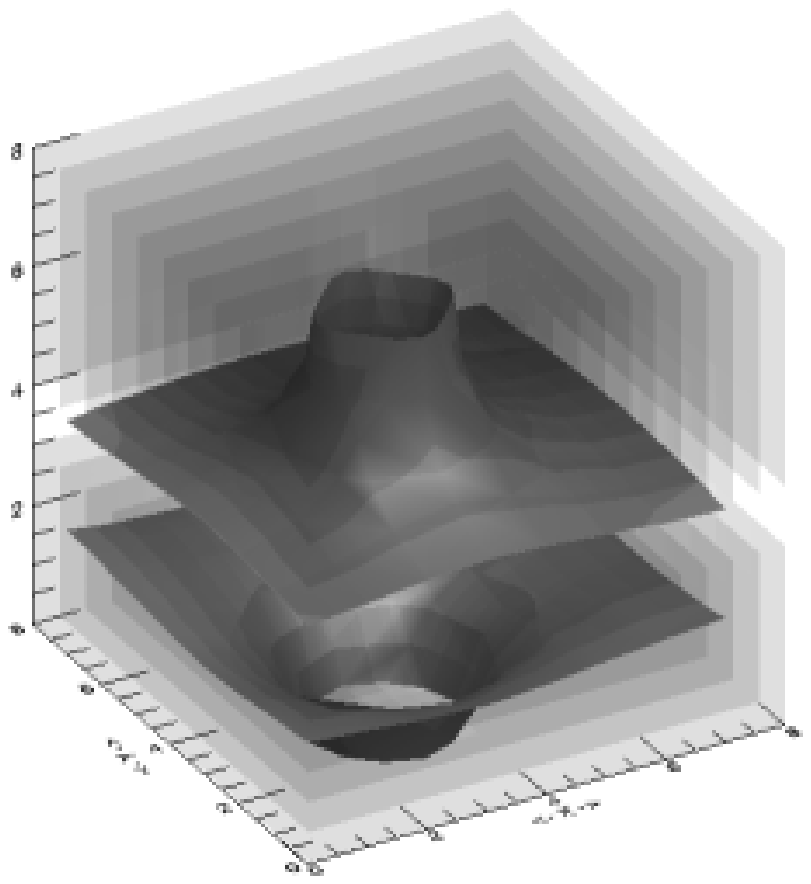,
	     bb=15 30 290 330, clip=, scale=0.25} \\                                 
\end{array}$
\end{center}          
\caption{\label{fig:artificialmode}The density shape for zero-mode eigenvector of
the constructed gauge configuration (lattice size $8^4$).
The 3D cuts are for slices $n_t=1\ldots 4$ (first row) and $n_t=5\ldots 8$ (second row). 
The inverse participation ratio for that mode is
$I=1.85$.}
\end{figure}


In 2D the Pontryagin index is constructed from $ \int d\sigma_{\mu\nu} F_{\mu\nu} $.
On the lattice for compact $U(1)$ gauge theory and with  torus topology various
explicit configurations with such topological charge have been explicitly constructed
as well as observed in Monte Carlo simulations (e.g. in Ref. \cite{GaHiLaGaHi}).  In 4D the
topological charge is proportional to $\int d^4x  F_{\mu\nu} \tilde{F}_{\mu\nu}$ and a
possible method \cite{Ba82,SmVi87a} to construct objects with non-vanishing
topological charge is e.g. to combine two-dimensional submanifolds, both with
topological charge in the 2D definition. Intersections of 2D surfaces in 4D may be
points as  well as more complicated, e.g. line-like objects. In Ref. \cite{BeHeMa01} such a
configuration - as suggested in Ref. \cite{SmVi87a} has been studied and identifies as
leading indeed to exact zero-modes. For unit topological charge (lattice size $L^4$,
$\omega=2\pi/L^2$) it has the form
\begin{eqnarray}
U_1(x)&=&\exp{(i\,\omega \,x_2)}\,,\\
U_2(x)&=&1 \quad\textrm{for~}x_2=1\ldots L-1\,\,, \\
&&\exp{(-i\,\omega \,L\, x_1)} \quad\textrm{for~}x_2=L\,,
\end{eqnarray}
and for $U_3$ and $U_4$ equivalent, with $x_3$ and $x_4$ replacing $x_1$ and $x_2$.
Fig.\ \ref{fig:artificialmode} shows the geometric shape of the eigenvector density of
this mode and we observe tubular structures. Note, however, that the tubes are in the
3D intersections, i.e. extend like planes in 4D.

\section{Conclusion} 

We have studied zero-mode properties of the overlap Dirac operator  for $U(1)$ gauge
theory in the confined phase. We find no correlation between the occurrence of these
modes and the monopole density confirming \cite{BeHeMa01}. 

We do observe individual zero-modes with definite chirality. This would contradict
the Atiyah-Singer  index theorem if applied naively for lattice systems. However,
for $U(1)$ lattice gauge theory for compact gauge fields the situation changes. In
particular the confinement phase of this system has no analytic continuation to the
weak coupling continuum limit. In the confinement phase one can explictly construct
configurations with topological charge and zero-modes \cite{SmVi87a}, essentially by
combining two 2D sub-manifolds with  non-vanishing 1st Chern number.

In the Coulomb phase there are no zero-modes and also the near-zero-modes are
suppressed. There we find no pronounced localization signal.  In the confinement
phase the exact zero-modes are definitely more localized than other eigenmodes (as
exhibited by the inverse participation ratio). The IPR of the nearby non-zero modes
does not seem to depend on the Im($\lambda$).  The geometrical structure of the zero
eigenmodes has no clear signature singling out 4D blobs - in contrast to e.g. the
case of instantons  in QCD, which have been observed in distorted hyperspherical
shapes. All evidence points to tubular or even planar structures supporting the
density function $p(x)$ for a zero-mode.


{\bf Acknowledgment:}
We want to thank Elizabeth\ Gasparim, Christof\ Gattringer and Pushan\ Majumdar
for discussions and Wolfgang Bietenholz for helpful comments.
Support by Fonds zur F\"orderung der Wissen\-schaft\-li\-chen Forschung in \"Osterreich,
project P16310-N08 is gratefully acknowledged.

\end{document}